\documentclass[twocolumn,amsmath,amssymb]{snp}
\usepackage{graphicx}
\usepackage{dcolumn}
\usepackage{bm}
\begin{document}

\title{{\Large Quark-Gluon Plasma: Present and Future }}

\bigskip
\bigskip
\author{\large Tapan K. Nayak\footnote{Dedicated to the memory of Prof. Aswini Kumar Rath who was the Local
Convenor of the DAE Symposium on Nuclear Physics (December 2007) at Sambalpur University where the talk was presented.}}
\affiliation{Variable Energy Cyclotron Centre, Kolkata - 700064, India}
\bigskip
\bigskip

\begin{abstract}
\leftskip1.0cm
\rightskip1.0cm
We review a sample of the experimental results from AGS to SPS and RHIC 
and their interpretations towards understanding of the Quark-Gluon Plasma.
We discuss extrapolations of these results to the upcoming LHC experiments.
Finally, we present the plans to probe the QCD critical point with an energy
scan at RHIC and FAIR facilities.
\end{abstract}

\maketitle

\bigskip
\bigskip
\noindent
{\large{\bf Introduction}}
\medskip

It is about thirty years since the intense study of hot and dense nuclear matter 
in the form of Quark-Gluon Plasma (QGP) has started. A tremendous amount of effort 
has gone in for the development of four generations of experiments and a large number 
of data have been collected and analyzed. Major theoretical developments have been 
made at the fundamental level and theoretical models have been developed in order to 
understand the implications of the data. A glimpse of ongoing activities may be found in \cite{qm2006}.
It is appropriate at this time to ask: 
(i) what were the expectations in the beginning? 
(ii) What have we learned so far? 
(iii) What are the prospects for the future?

Based on the two most novel properties of QCD, {\it viz}, asymptotic freedom of 
quarks and quark confinement, it was conjectured that it would be worthwhile  
to explore these phenomena by creating high energy and high density matter over a 
large volume. At these high temperatures and densities normal nuclear matter is
expected to undergo a phase transition to a new state of matter, called the 
Quark-Gluon plasma. 
Interesting connections were made to the big bang model of the early 
stages of our Universe where a QGP state might have existed. 
Intense experimental programs with collisions of 
nuclei at relativistic energies have started to recreate these conditions in the laboratory. 

The quest for the search and study of QGP started in early eighties 
with the acceleration of
Au beam at 1~GeV/A at the Bevalac. The early success of the experiments in terms
of bringing out the collective nature of the matter produced prompted the scientists
at Brookhaven National Laboratory (BNL) 
and CERN to make concrete programs for the future
accelerator developments for heavy ions. The next milestone came with the acceleration
of Au beam at 11.7~GeV/A at the AGS at BNL and Pb beam at 158~GeV/A at the CERN-SPS.
First hints of the formation of a new state of matter has been obtained from
the SPS data in terms of global observables, event-by-event fluctuations, direct photons, di-leptons and
most importantly, the $J/\psi$ suppression. The Relativistic Heavy Ion Collider (RHIC) started becoming operational
in the year 2000 with \mbox{Au-Au} collisions at $\sqrt{s_{\rm NN}}=130$~GeV and
soon after to top  \mbox{Au-Au} energies of $\sqrt{s_{\rm NN}}=200$~GeV. The
experimental program at RHIC included four experiments, two large and two small with
the involvement of more than 1200 physicists. At present, the RHIC experiments bring 
out highest quality data from \mbox{p-p}, \mbox{Cu-Cu} and \mbox{Au-Au} at various
energies. Strong evidence for the production of extreme hot and dense matter has
been seen. The matter formed at RHIC has been termed as 
sQGP (strongly coupled QGP). The
RHIC results, in combination with the ones from AGS and SPS, have enhanced our
understandings of the QCD matter at different temperatures and densities.

The CERN large hadron collider (LHC) is expected to be commissioned within one year.
Heavy-ion physics is an integral part of the baseline program of the LHC which will
accelerate \mbox{Pb-Pb} beams at $\sqrt{s_{\rm NN}}=5.5$~TeV, a factor 27 higher than the 
top RHIC energy. This increase is even larger than the factor 10 in going from the 
CERN-SPS to RHIC. It will lead to significant extension of the kinematic range in transverse 
momentum ($p_{\rm T}$) and in Bjorken $x$. LHC will turn out to be a discovery machine
for various types of physics and will explore QCD phenomena in great detail.

\begin{figure}
\includegraphics[scale=0.97]{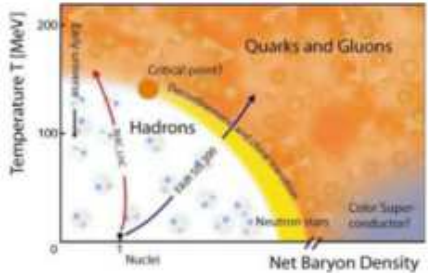}
\caption{\label{fig:phase_diagram} Phase diagram of nuclear matter.}
\end{figure}

The QCD phase diagram, 
as shown in Figure \ref{fig:phase_diagram}, characterized
by the temperature (T) and the baryon chemical potential $\mu_{\rm B}$, signifies the
separation of QGP to hadronic phase. 
One of the major predictions of QCD in extreme conditions of high
temperature or large baryon number density is the existence of a critical point
at a particular temperature and density where a sharp transition between the QGP 
phase and the hadronic phase first appears. 
It may be possible to access the critical point 
experimentally by scanning the QCD phase diagram
in terms of T and $\mu_{\rm B}$. This can be accomplished 
by varying beam energies 
from about $\sqrt{s_{\rm NN}}$=5~GeV to 100~GeV. 
Such a program has recently been undertaken
at RHIC\cite{rhic}. Experiments at GSI\cite{CBM} are planned to study this as well.
The discovery of the critical point would be very important to the QGP study.

In this review, we present selected experimental results from
AGS to SPS to RHIC and predictions for the LHC. At the end we will return back
to the questions of where we are and what to expect next.

\begin{figure}
\includegraphics[scale=0.55]{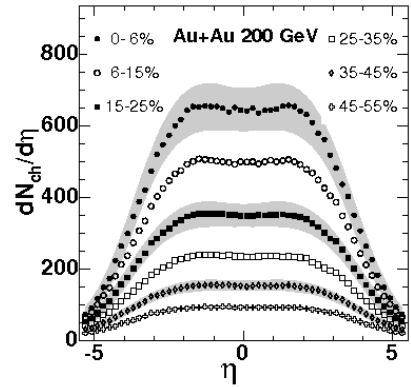}
\caption{\label{fig:phobos_dedeta} Charged particle multiplicity density at RHIC
energy for different collision centralities \cite{phobos}.   }
\end{figure}

\bigskip
\noindent
{\large{\bf Global observables }}
\medskip

\begin{figure}
\includegraphics[scale=0.45]{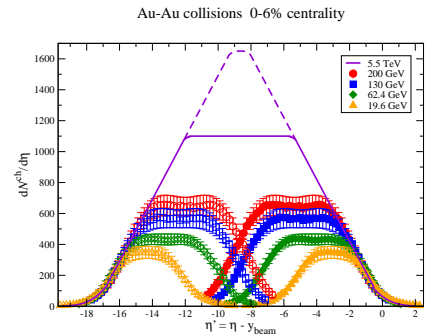}
\caption{\label{fig:lhc_dndeta} Charged particle multiplicity density scaled by their
beam rapidities and extrapolation to LHC energies \cite{urs}. }
\end{figure}

The comprehensive study of particle production, rapidity distributions,
particle ratios, momentum spectra, flow and source size estimations provide valuable information
for thermal and chemical analysis of the freeze-out conditions. One of
the first results which came out from RHIC is shown in
Figure \ref{fig:phobos_dedeta} in terms charged particle multiplicity
density for \mbox{Au-Au} collisions at different collision centralities \cite{phobos}.
Combining all the data from SPS to RHIC, one can make an extrapolation
to LHC \cite{urs} as shown in Figure \ref{fig:lhc_dndeta}. These data
are plotted in the rest frame of one of the colliding nuclei (full symbols) 
and mirrored at LHC mid-rapidity (open symbols). At LHC energies, 
a pseudorapidity density close to 1100 at the mid rapidity is obtained. Other estimations
for LHC energies give numbers between 1200-2500 for pseudorapidity density at
mid rapidity \cite{urs}. 

\begin{figure}
\includegraphics[scale=0.4]{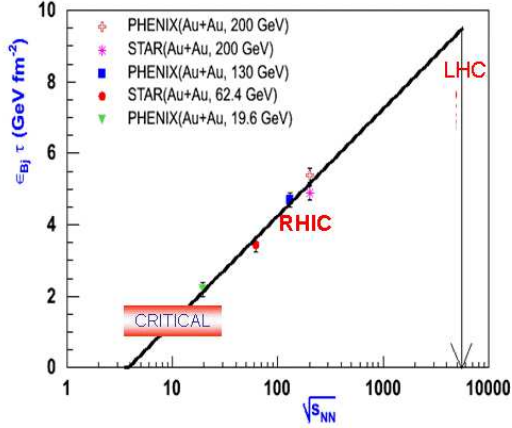}
\caption{\label{fig:Bjorken} Energy density as a function of beam energy. The figure
indicates the possible location of the critical point and 
extrapolation to the LHC energy.}
\end{figure}

Energy density estimations have been made using the rapidity densities
and mean transverse momenta. The values for RHIC energies are shown in
Figure \ref{fig:Bjorken} and extrapolated to LHC energy \cite{raghu}. The energy
density where the critical point occurs is shown in this figure. The energy
density achieved at RHIC energies are already seen to be beyond the critical density
for QGP formation.

\begin{figure}
\includegraphics[scale=0.5]{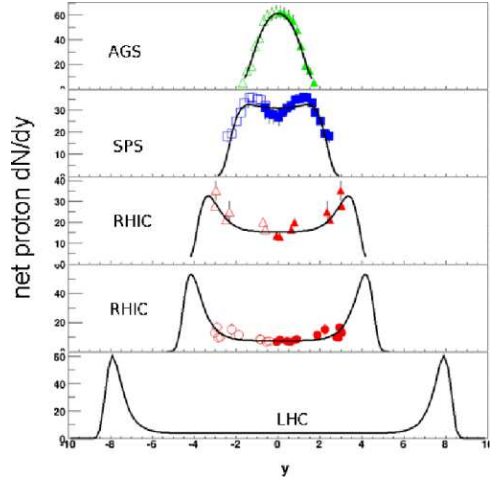}
\caption{\label{fig:net_proton} Net proton rapidity distributions at various 
collision energies.}
\end{figure}

The measured net proton rapidity density distributions for AGS, SPS, RHIC energies 
\cite{esumi}
are shown in Figure \ref{fig:net_proton} with extrapolations to LHC energies. The figure
shows that a complete transparency can be expected at LHC energies for a large rapidity range.

\begin{figure}
\includegraphics[scale=0.7]{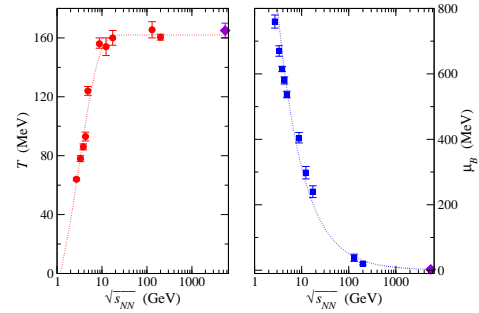}
\caption{\label{fig:temp_mu}Estimation of freeze-out temperature and chemical
potential from thermal model fits as a function of centre-of-mass energy of the
collision \cite{urs}.  }
\end{figure}

\begin{figure}
\includegraphics[scale=0.7]{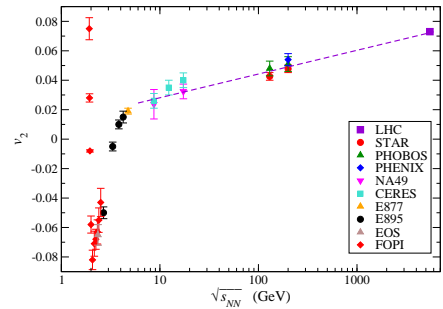}
\caption{\label{fig:lhc_v2} 
Elliptic flow, $v_2$ as a function of centre-of-mass energy of the
collision \cite{urs} for existing data and extrapolated to LHC.
}
\end{figure}

From the measured spectra and particle ratios, it is possible to estimate the 
freeze-out temperature and chemical potential by using thermal model fits \cite{urs}.
This is shown in Figure~\ref{fig:temp_mu} which can be used to map the QCD phase space. The
chemical potential at top RHIC energies is between 20-40MeV and at LHC energies it
is expected to be less than 10MeV. Since various lattice calculations give different
values of chemical potentials (180MeV  to 500MeV) for the critical point, it is
obvious that a thorough energy scan is needed from $\sqrt{s_{\rm NN}}$=5GeV to 100GeV,
in order to probe the region around the critical point.

An important measure of the collective dynamics of heavy-ion collisions is the elliptic
flow ($v_2$). Figure~\ref{fig:lhc_v2} shows excitation function of $v_2$ for mid-central
collisions. Because of the large values of $v_2$ at RHIC energies, in agreement with
the value for an ideal fluid, the formation of a perfect liquid is ascertained 
at RHIC energies.

The dynamical evolution of the collision fireball and its
space-time structure has been traditionally studied using two-particle (HBT)
correlations. The multiplicity and transverse momentum dependence of three-dimensional 
pion interferometric radii in Au-Au and Cu-Cu collisions at different RHIC
energies \cite{das} have been shown in Figure~{\ref{fig:hbt}.
The freezeout volume estimates with charged pions measured from such studies,
show linear dependence as a function of charge particle multiplicity
indicating consistent behaviors with a universal mean-free-path at freeze-out.
The HBT correlations of photons is expected to provide better insight into
the nature of the evolving system \cite{bass_photon_hbt,wa98_photon_hbt,das_photon_hbt}.
The HBT correlation studies can be studied in more detail at the LHC.
Moreover, it may be possible to extract HBT radii on an event by event basis 
at the LHC energies.

\begin{figure}
\includegraphics[scale=0.31]{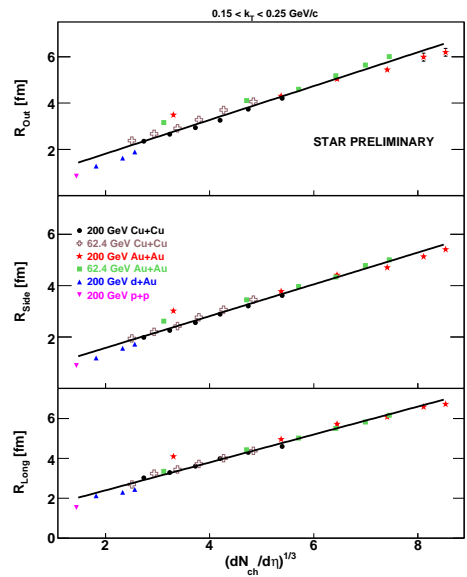}
\caption{\label{fig:hbt} Pion source radii dependence on charged particle
multiplicity. The lines are plotted to guide the eye and represent
linear fits to the data.}
\end{figure}

\bigskip
\noindent
{\large{\bf Event-by-event Fluctuation}}
\medskip

Fluctuations of thermodynamic quantities provide an unique
framework for studying the nature of the phase transition and provide direct insight
into the properties of the system created \cite{nayak}. Large fluctuations in energy density
due to droplet formation are expected if the phase transition is of first order and a second order phase transition might lead to divergence in specific heat and increase in the fluctuations. 
Fluctuations are also predicted to be largely enhanced near the critical point. Fluctuations
have normally been studied in terms of $<p_{\rm T}>$ and temperature, multiplicity, strangeness,
net-charge, balance functions, azimuthal anisotropy and source sizes. The formation
of disoriented chiral condensates is expected to lead to large charged-neutral fluctuations.

Relative production of different particle species produced in the hot and dense matter
might get affected when the system goes through a phase transition.
Large broadening in the ratio of kaons to pions
has long been predicted because of the differences in free enthalpy of the
hadronic and QGP phase. This could be probed through the fluctuation in
the $K/\pi$ ratio. The dynamic fluctuation, $\sigma_{\rm dyn}$, in the $K/\pi$ ratio 
(Figure~\ref{fig:star_kpi}) is seen to decrease with beam energy in going from AGS to SPS energies
and then remain constant. 

\begin{figure}
\includegraphics[scale=0.64]{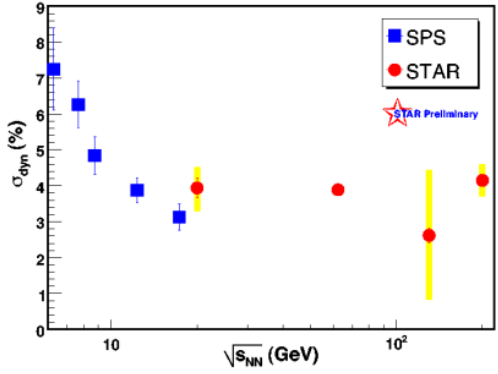}
\caption{\label{fig:star_kpi}Excitation function for $\sigma_{dyn}$ of 
$[K^{+}+K^{-}]/[\pi^{+}+\pi^{-}]$ ratio at the SPS (left panel) and
with an extension to RHIC (right panel) \cite{star_kpi} }
\end{figure}

\begin{figure}
\includegraphics[scale=0.55]{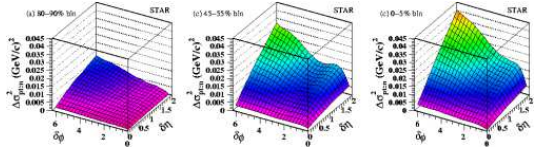}
\caption{\label{fig:pt_etaphi}Scale dependence of $<p_{\rm T}>$ fluctuation within
the STAR acceptance expressed in terms of per-particle variance difference 
\cite{star_pt_etaphi}. Results are given for three collision centralities (from peripheral to
central in going from left to right). }
\end{figure}

In order to be
more sensitive to the origin of fluctuations, differential
measures have been adopted where the analysis is performed at
different scales (varying bins
in $\eta$ and $\phi$). The scale dependence of $<{p_{\rm T}}>$ fluctuation for
three centralities in \mbox{Au-Au} collisions at
$\sqrt{s_{\rm NN}}=200$~GeV \cite{star_pt_etaphi} is shown 
in Figure~\ref{fig:pt_etaphi}. The extracted auto correlations are
seen to vary rapidly with collision centrality, suggesting that
fragmentation is strongly modified by a dissipative medium in
more central collisions relative to peripheral collisions. Further
studies for different charge combinations will provide more
detailed information.

\bigskip
\bigskip
\noindent
{\large{\bf High p$_T$ and jets }}
\medskip

\begin{figure}
\includegraphics[scale=0.43]{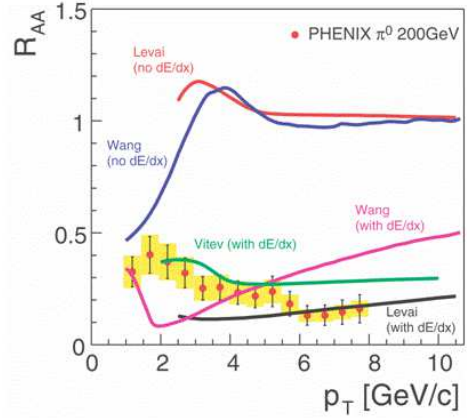}
\caption{\label{fig:RAAphenix} The nuclear modification factor $R_{\rm AA}$ for $\pi^0$ compared
to different model predictions for neutral pions \cite{RAAphenix}. }
\end{figure}

\begin{figure}
\includegraphics[scale=0.4]{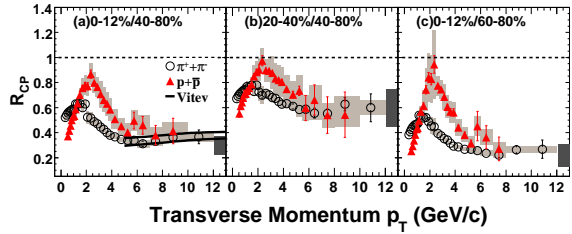}
\caption{\label{fig:RAAstar} The nuclear modification factor $R_{\rm CP}$ for identified charged 
pions and protons \cite{RAAstar}.
}
\end{figure}

Properties of the hot and dense medium produced in nucleus-nucleus collisions can be 
studied via the energy loss experienced by fast partons in the medium. Detailed measurements
have been performed by experiments at RHIC for the nuclear modification factor, $R_{\rm AA}$,
defined as,
\begin{eqnarray}
  R_{\rm AA} = \frac{dN_{Au+Au}}{<T_{AA}>d\sigma_{p+p}}, \nonumber  
\end{eqnarray}
for neutral and identified particles. Figures \ref{fig:RAAphenix} and \ref{fig:RAAstar} give 
$R_{\rm AA}$ and $R_{\rm CP}$ (defined as ratio of central to peripheral collisions)
as function of $p{\rm T}$ as measured by the PHENIX \cite{RAAphenix} and STAR 
experiments \cite{RAAstar}, respectively. Strong suppressions are seen in central \mbox{Au-Au}
collisions corresponding to the \mbox{p-p}.
The results indicate that at low $p_{\rm T}$, protons and anti-protons are less
suppressed than charged pions whereas at high $p_{\rm T}$ the partonic sources of pions and protons have
similar energy loss when traversing the nuclear medium. More detailed results may be found in
 \cite{RAAphenix} and \cite{RAAstar}.

\begin{figure}
\includegraphics[scale=0.35]{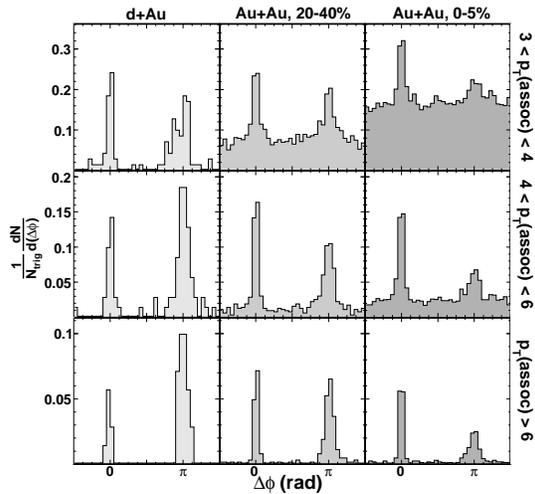}
\caption{\label{fig:star_dijet} Azimuthal correlation of high $p_{\rm T}$ charged hadrons.
\cite{star_dijet}.}
\end{figure}

The back-to-back correlation strength of high $p_{\rm T}$ hadrons is seen to be
sensitive to the in-medium path length of the parton. The study by the STAR 
experiment over a broad range in transverse momenta \cite{star_dijet} is shown in 
Figure~\ref{fig:star_dijet} for \mbox{d-Au} and  \mbox{Au-Au} collision at two
different centralities. Whereas there is no significant change in the near side
peak with the increase of $p_{\rm T}$ of the associated particles, 
the away-side correlation strength decreases 
from \mbox{d-Au} to central \mbox{Au-Au} collisions. 
The strongest modifications of the correlated yields are seen at lower associated $p_{\rm T}$. 
More detailed measurements of nuclear modification factors and azimuthal asymmetries should
be made in order to constrain the theoretical models.

The measurement of $\gamma$-jet events provides an unique probe of parton energy 
loss \cite{renk}. These studies are being actively pursued by RHIC experiments \cite{marco}.

\bigskip
\noindent
{\large{\bf Electromagnetic probes and quarkonia }}
\medskip

\begin{figure}
\includegraphics[scale=0.45]{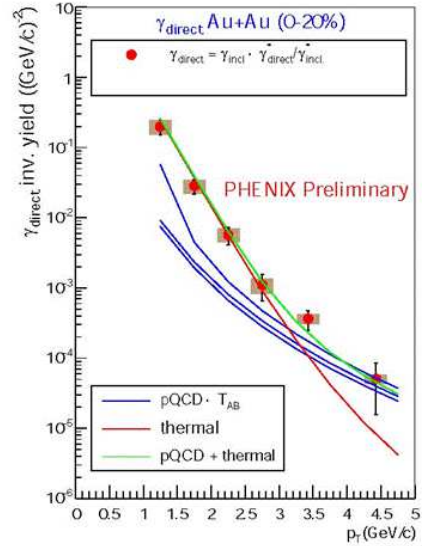}
\caption{\label{fig:photon_phenix} Yield of direct photons as a function of
$p_{\rm T}$ for central \mbox{Au-Au} collisions compared to various theoretical models.}
\end{figure}

Electromagnetic probes, {\t viz.}, photons and dileptons have long been recognized as the
most direct probes of the collision system. Owing to the nature of the interaction they undergo
minimal scatterings and are by far the best markers of the entire space-time evolution
of the collision. The single photon data obtained from \mbox{Pb-Pb} collisions at CERN-SPS by the 
WA98 Collaboration have been the focus of considerable interest \cite{wa98-photon}. The direct
photon spectra at low $p_T$ from the PHENIX experiment \cite{photon_phenix}
is shown in \ref{fig:photon_phenix}. The yield is consistent with rates calculated with 
thermal photon emission taken into account.

\begin{figure}
\includegraphics[scale=0.45]{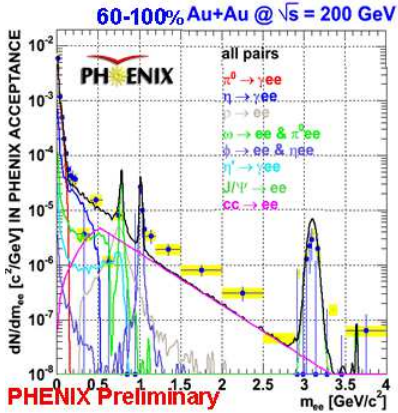}
\caption{\label{fig:dielectron_phenix} The dielectron spectra \mbox{Au-Au} collisions at 
$\sqrt{s_{\rm NN}}=200$~GeV together with cocktail hadron decay
sources and a pythia calculation of charm decays \cite{averbeck}}
\end{figure}

The dielectron spectra have been measured by the PHENIX
experiment \cite{averbeck}. This is shown in Figure~\ref{fig:dielectron_phenix} 
after background subtraction for  \mbox{Au-Au} collisions at 
$\sqrt{s_{\rm NN}}=200$~GeV together with cocktail hadron decay
sources and a pythia calculation of charm decays. The data are in good agreement with the
cocktail over the full mass range. Improvement in the data quality with respect to
the combinatorial background and better theoretical calculations are needed in order
to make any conclusive statement.

\begin{figure}
\includegraphics[scale=0.4]{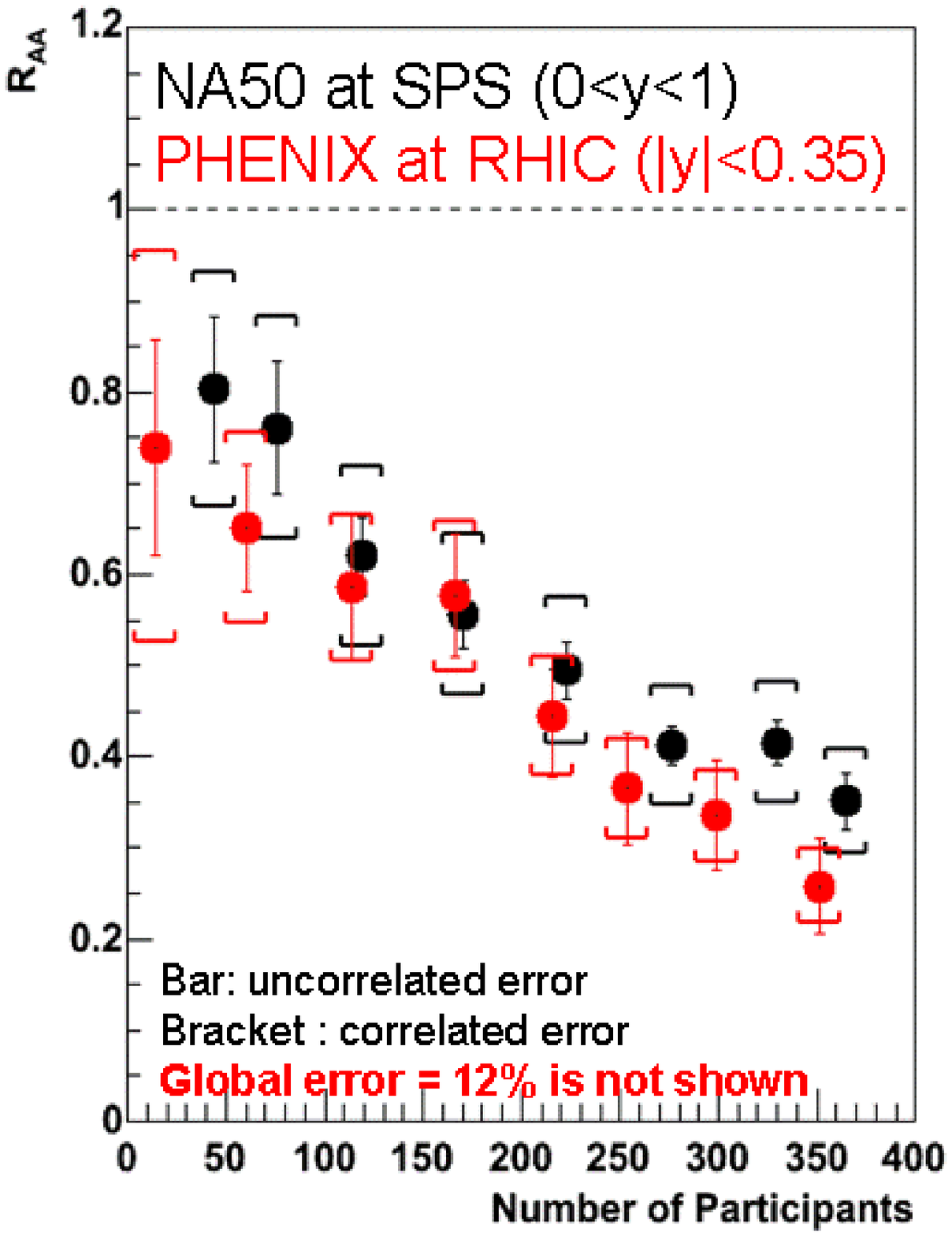}
\caption{\label{fig:j-psi} }
Nuclear modification factor for $J/\Psi$ for PHENIX results compared to those
from NA50.
\end{figure}

$J/\Psi$ suppression is considered to be one of the most direct signatures of QGP 
formation \cite{matsui}. 
The observed suppression of $J/\Psi$ in \mbox{Pb-Pb} collisions at the SPS energies
\cite{NA50} was considered to be a direct observation of deconfined matter.
However the above statement has been contrasted by several theoretical calculations.
Recent high accuracy measurement by the NA60 collaboration for \mbox{In-In} collisions
shows that a suppression is also present in these collisions \cite{NA60}.
In the meantime new results from the PHENIX experiment \cite{jpsiphenix} have been
compared to NA50 results which show reasonably good agreement. Higher statistics
data from PHENIX and STAR are necessary for distinctions between model predictions.

\bigskip
\noindent
{\large{\bf QCD at small $x$ and forward physics }}
\medskip

\begin{figure}
\includegraphics[scale=0.61]{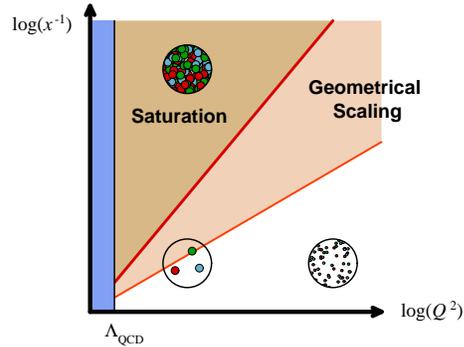}
\caption{\label{fig:gelis1} Saturation domain in the ($Q^2,x$) plane.}
\end{figure}

At large collision energy and relatively low momentum transfer ($Q$), one expects a new
regime of QCD known as saturation \cite{gelis}. This is described in a picture of
colour glass condensate (CGC) where a saturation scale emerges naturally. This is
pictorially depicted in Figure~\ref{fig:gelis3} in terms of the saturation domain.

\begin{figure}
\includegraphics[scale=0.97]{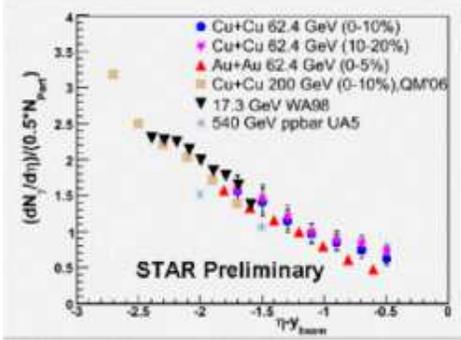}
\caption{\label{fig:star-pmd} Limiting fragmentation as observed from the rapidity 
density distribution for different collision energies and collision systems \cite{star-pmd1,star-pmd2}. }
\end{figure}

\begin{figure}
\includegraphics[scale=0.4]{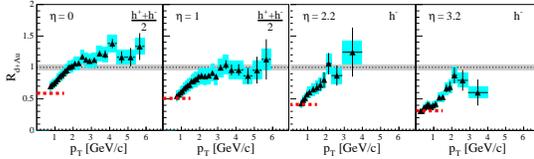}
\caption{\label{fig:gelis3} Forward suppression observed by BRAHMS for d+Au 
collisions at RHIC \cite{bearden}.}
\end{figure}

One of the experimental results which support the saturation phenomenon is the limiting
fragmentation \cite{star-pmd1,star-pmd2}, shown in Figure~\ref{fig:star-pmd} for several beam energies
and colliding systems. By shifting the rapidity axis by the beam rapidity, one
can see that the rapidity distribution for produced particles at collisions of various 
energies and system tend to an universal curve in the fragmentation region.
This property naturally follows from the CGC framework. The second observation is seen from the
suppression of hadron spectra at forward rapidity in \mbox{d-Au} collisions \cite{bearden}
from the BRAHMS experiment at RHIC as shown in Figure~\ref{fig:gelis3}. The suppression
of the nuclear modification factor at forward rapidities is considered to be a consequence of
the shadowing that builds up via the evolution in rapidity.

The ALICE experiment at LHC \cite{alice-ppr-1,alice-ppr-2} will probe a 
continuous range of $x$ as low as about 10$^{-5}$, accessing 
a novel regime where strong nuclear gluon shadowing is expected. 
The study of low $x$ regime, especially at forward rapidities, will be
most appropriate to study the early stage of nuclear collision.


\bigskip
\noindent
{\large{\bf Where are we now and what to expect}}
\medskip

The field of Quark-Gluon Plasma has reached a very interesting point of its lifetime.
We have the wealth of results from AGS, SPS and RHIC with us.
The theoretical developments have concentrated on a complete description
of the interactions of hadrons at high energy and their subsequent evolution 
into a thermalized quark-gluon plasma. There has been indication of the formation of a 
coexisting phase of quarks and hadrons 
even at SPS energies \cite{bedanga_van_hove}.
The matter formed at
RHIC is observed to be of low viscosity and high opacity, and is termed as
strongly coupled quark-gluon plasma (sQGP). 
There are experimental evidences for the colour glass condensate picture which 
considers the high density gluonic matter. More detailed studies, with high
$p_{\rm T}$ probes and high statistics data from SPS and RHIC, are being made to understand
the properties of the high temperature and high dense matter. With the
advent of very high energy beams at the LHC, all the three major
experiments (ALICE, CMS and ATLAS) are gearing up to study the new form
of matter in great detail \cite{LHC_predictions}. 

The QCD phase boundary is slowly getting mapped with 
data points from various experiments.
One important point which is missing from our experimental radar 
is the QCD critical point. We need to
have a good guidance from the lattice
calculations with regard to the location of the critical point. The exact location
of the critical point is not known yet. A new program for
the RHIC low energy scan to search for the QCD critical point is underway
which will also cover a broad region of physics interest \cite{rhic}. The STAR experiment
is expected to take a lead in this regard. The CBM experiment \cite{CBM} in the newly
planned FAIR facility at GSI will be able to probe the critical point and
study it in detail. 

We would like to think that the field of Quark-Gluon Plasma is 
very much in its youth and we
expect a lot of fundamental and interesting physics to happen within the next decade.

\bigskip
\noindent
{\large{\bf Acknowledgments}}
\medskip

I would like to thank the Organizers of the DAE Nuclear Physics Symposium to give me
the opportunity to review the field of Quark Gluon Plasma. I would like
to acknowledge the contribution of the members of the
Indian Collaboration who are actively involved
in the QGP study at BNL and CERN.

\end{document}